\documentclass[conference]{IEEEtran}
\IEEEoverridecommandlockouts

\usepackage[cmyk]{xcolor}
\usepackage[hidelinks]{hyperref}
\usepackage[utf8]{inputenc}
\usepackage{newtxtext,newtxmath}
\usepackage[american]{babel}
\usepackage{csquotes}
\usepackage[final]{microtype}
\usepackage{multirow}
\usepackage{xurl}
\usepackage{booktabs}
\usepackage{tabularx,ragged2e}
\newcolumntype{L}{>{\raggedright\arraybackslash}X}
\newcolumntype{R}{>{\raggedleft\arraybackslash}X}
\usepackage{threeparttable}
\usepackage{stmaryrd}
\usepackage{float}
\usepackage{adjustbox}

\usepackage{pgfplots}
\pgfplotsset{compat=1.17}
\usepackage[outline]{contour}
\contourlength{0.2em}
\usepackage{tikz}
\usetikzlibrary{patterns}
\usepgfplotslibrary{fillbetween}
\usetikzlibrary{patterns}
\usetikzlibrary{positioning, calc}

\usepackage{siunitx}
\DeclareSIUnit\tx{tx}
\DeclareSIPrefix{\noop}{}{0} % https://tex.stackexchange.com/a/605736/47127
\usepackage[backend=biber, style=ieee]{biblatex}
\usepackage{acro}
\DeclareAcronym{PoW}{
  short = PoW,
  long  = proof-of-work
}
\DeclareAcronym{PoS}{
  short = PoS,
  long  = proof-of-stake
}
\DeclareAcronym{DLT}{
  short = DLT,
  long  = distributed ledger technology
}
\DeclareAcronym{TPS}{
  short = TPS,
  long  = transactions per second
}
\DeclareAcronym{GHG}{
  short = GHG,
  long  = greenhouse gas
}
\DeclareAcronym{CO2}{
  short = CO2,
  long  = carbon dioxide
}
\DeclareAcronym{GJ}{
  short = GJ,
  long  = Giga joules
}

\usepackage[nomessages]{fp}

\newif\ifdraft
\drafttrue
\ifdraft
\newcommand{\jinote}[1]{ {\textcolor{purple} { ***Juan: #1 }}}
\newcommand{\afnote}[1]{ {\textcolor{green} { ***Alex: #1 }}}

\else
\newcommand{\jinote}[1]{}
\newcommand{\afnote}[1]{}
\fi

\DeclareRobustCommand*{\IEEEauthorrefmark}[1]{%
  \raisebox{0pt}[0pt][0pt]{\textsuperscript{\footnotesize #1}}%
}

\begin{document}

\title{The energy consumption of Proof-of-Stake systems: Replication and expansion}

\author{
    \IEEEauthorblockN{
        Juan Ignacio Ibañez\IEEEauthorrefmark{1}\IEEEauthorrefmark{2}\IEEEauthorrefmark{3},
        Francisco Rua\IEEEauthorrefmark{1}
    }

    \IEEEauthorblockA{\IEEEauthorrefmark{1}Centre for Blockchain Technologies, University College London, London, UK}
    \IEEEauthorblockA{\IEEEauthorrefmark{2}DLT Science Foundation}
    \IEEEauthorblockA{\IEEEauthorrefmark{3}Facultad de Ciencia Política y Relaciones Internacionales, Universidad Católica de Córdoba, Córdoba, Argentina}
    \IEEEauthorblockA{j.ibanez@ucl.ac.uk, francisco.rua@mi.unc.edu.ar
}
}

\maketitle

% 150 Word Limit!
\begin{abstract}
Blockchain technology and, more generally, \ac{DLT} systems, face public scrutiny for their energy consumption levels. However, many point out that high energy consumption is a feature of (small block size) \ac{PoW} \ac{DLT}s, but not of \ac{PoS} \ac{DLT}s. With the energy consumption of \ac{PoS} systems being an under-researched area, we replicate, expand and update embryonary work modelling it and comparing different \ac{PoS}-based \ac{DLT}s with each other and with other non-\ac{PoS} systems. In doing so, we suggest and implement a number of improvements to an existing \ac{PoS} energy consumption model. We find that there may be significant differences in the energy consumption of \ac{PoS} systems analysed and confirm that, regardless of these differences, their energy consumption is several orders of magnitude below that of Bitcoin Core.
\end{abstract}

% 5 Keyword Limit!
% Avoid keywords that appear in the title to increase the probability of being found
\begin{IEEEkeywords}
Energy consumption, Blockchain, DLT, Sustainability, Proof of Stake, Technological Innovation.
\end{IEEEkeywords}

% Introduce problem, outline solution; the statement of the problem should include a clear statement why the problem is important (or interesting).
\section{Introduction}
\label{sec:introduction}

\ac{DLT} has positioned itself as one of the main technologies challenging the current technological landscape. \ac{DLT}, usually in the form of blockchain technology, is as innovative as it is controversial. As the rate of adoption of \ac{DLT} and crypto assets increases \cite{Hammond2021Cryptos2022}, so does the inquiry into its potential drawbacks and risks.

One of the issue areas increasingly under the public eye is the environmental impact of DLT, specifically through \ac{GHG} emissions that may aggravate climate change. This concern is echoed by the The White House Office of Science and Technology Policy, that recently released a report \cite{OSTP2022ClimateStates} seeking to summarise the state of the art in this regard. This is a part of the broader process triggered by the United States President's Executive Order 14067: “Ensuring Responsible Development of Digital Assets,” which further illustrates this concern.

The discussion is characterised by arguably three dimensions. First, the debate around whether \ac{PoW} mining should be considered environmentally harmful or environmentally friendly due to its energy-intensiveness or its non-rival consumption of energy, respectively \cite{OSTP2022ClimateStates,Ibanez2023CanSoK}. Second, the interest in \ac{PoS} as an alternative DLT consensus mechanism requiring significantly less energy consumption \cite{Platt2021TheProof-of-Work}. Third, the potential of \ac{DLT} to enable distributed energy resources and environmental markets \cite{OSTP2022ClimateStates,Mollah2021BlockchainSurvey}.

This article focuses on the second dimension, i.e. the characteristics of \ac{PoS} as an alternative consensus mechanism. The consensus mechanism in a \ac{DLT} performs several tasks, with the prevention of Sybil attacks\footnote{A sybil attack is a scenario in which “a node illegitimately claims multiple identities” \cite{Patel2017AWSN}, pretending “to be many nodes simultaneously by using many different addresses while transmitting” \cite{Piro2006DetectingNetworks}. In other words, it is the “forging of multiple identities (...) [to] control a substantial fraction of the system” \cite{Douceur2002TheAttack} (see also \cite{Trifa2014SybilAttack})} \cite{Platt2021TheProof-of-Work} being one of the most important. These attacks occur due to the low cost of creating identities in digital systems (ibid). \ac{DLT}s typically solve this issue by attaching the ability to influence the state of the network to a scarce resource. In \ac{PoW}, a validator (“miner”) needs to spend energy (“work”) to find a number (“nonce”) by trying as many hashes at random as possible (“mining”). The ability to influence the network state is proportional to the work spent in it, not to the number of identities controlled by the miner, hence solving the Sybil attack. Nevertheless, this is an energy-intensive process. Although this very feature is by design, it has led to criticism \cite{OSTP2022ClimateStates,Ibanez2023CanSoK}.

\ac{PoS}, instead, replaces the scarce resource of energy with that of “stake,” thereby eliminating mining itself as an activity. Because the deliberate energy expenditures of mining are what originates most of \ac{PoW}-based \ac{DLT}'s energy consumption, the consumption of \ac{PoS}-based \ac{DLT}s is considered minuscule in comparison. However, the exact extent to which this reduction goes, as well as the differences in this regard between the various \ac{PoS}-based \ac{DLT}s in the market, remains an under-studied area \cite{Platt2021TheProof-of-Work,Gallersdorfer2022EnergyProtocols,CCRI2022EnergyBlockchain,CCRI2022DeterminingNetworks}.

\section{Previous work}
% Previous or obvious approach

In 2021, Platt et al \cite{Platt2021TheProof-of-Work} began to fill this gap by analysing the energy consumption of leading \ac{PoS} \ac{DLT}s, namely Algorand, Cardano, Ethereum, Hedera, Polkadot and Tezos. The paper furthermore compared their energy consumption to that of Bitcoin Core as well as VisaNet.

The paper provides a useful contribution to the literature on many grounds. It provides a first exploration, it sets up a benchmark basic methodology for future studies, it considers how meaningful energy consumption comparisons should be considered not just in global terms nor just on a per transaction basis but also controlling by throughput, and it includes several of the leading \ac{PoS}-based \ac{DLT}s.

The authors developed a simple method to quantify the possible energy consumption per transaction (kWh/tx) of a PoS-based \ac{DLT}. In short, it consists in taking observations of the throughput and number of validators at different points in time and using this data to extrapolate a function where the number of validators (and, indirectly, their energy consumption) is a function of throughput. They resort to different assumptions for the energy consumption per validator to arrive at a range of possible energy consumption values for each throughput value per PoS-based \ac{DLT} (with the function’s domain being established by the \ac{DLT}’s energy consumption at maximum throughput). They supplement this with the upper and lower thresholds of Bitcoin’s electricity consumption per transaction according to the  Cambridge Bitcoin Electricity Consumption Index \cite{CBECI2023CambridgeIndex}; and with VisaNet’s energy consumption and the average of transactions per day according to its annual ESG Report.

Because the Platt et al \cite{Platt2021TheProof-of-Work} paper pre-dates the Ethereum Merge, no actual real-world measurements of Ethereum’s (then called “Ethereum 2.0”) throughput and energy consumption were inputted in this model. The authors instead resorted to Ethereum’s pre-Merge throughput and to ex-ante estimations of the lower and upper bounds of Ethereum’s post-Merge energy consumption.

\begin{figure*}[h]
\centering
\includegraphics[width=\textwidth]{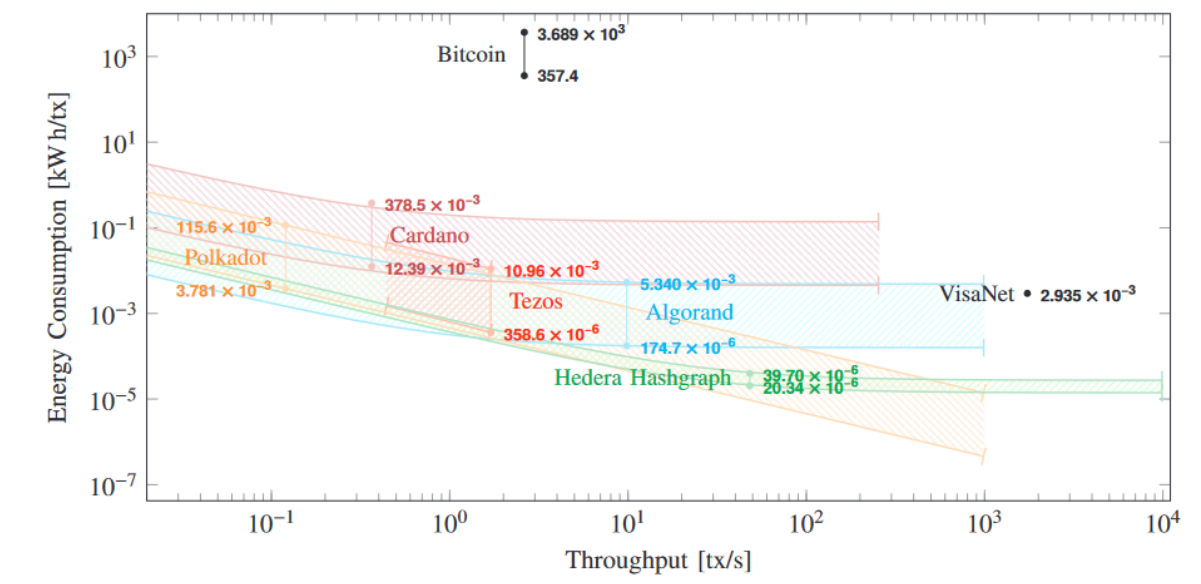}
\caption{Throughput and energy consumption per transaction in Platt et al \cite[1140]{Platt2021TheProof-of-Work}. The lower boundary for each curve indicates the energy consumption under an optimistic validator hardware assumption while the upper boundary indicates a pessimistic scenario. The two points per \ac{DLT} indicate the latest observation.}
\label{figure:platt}
\end{figure*}

One of the main outputs of the aforementioned paper is reproduced in Figure \ref{figure:platt}. The figure shows the comparison of the several \ac{PoS}-based \ac{DLT}s in terms of energy consumption per transaction for different levels of throughput. It suggests that the energy consumption of different \ac{PoS}-based \ac{DLT}s may differ substantially, that the Bitcoin network offers a high energetic cost in contrast to PoS-based alternatives, and that distributed options may compete with centralised systems such as VisaNet also on energy terms.

\section{Replication and expansion}
% Approach/solution/contribution

This article seeks to replicate, expand and update the work laid out by Platt et al \cite{Platt2021TheProof-of-Work}. The reasons for this are intuitive: replication is a fundamental aspect of the scientific process, two important blockchains had not been included in Platt et al \cite{Platt2021TheProof-of-Work}, Platt et al’s paper was supported but a very limited number of observations, and there was no real-world data for post-Merge Ethereum in 2021.

The current article is thus designed as follows:

\begin{enumerate}
    \item The universe of \ac{PoS}-based \ac{DLT}s studied includes Algorand, Cardano, Ethereum, Hedera, Polkadot and Tezos, but also Avalanche, BNB Chain, Elrond/MultiversX, Flow, NEAR, Toncoin, Tron, and Solana. This article hence captures the top 50 PoS-based layer-1 (L1) \ac{DLT}s by market cap at the time of release of this working paper.
    \item Real-world observations are taken for post-Merge Ethereum.\footnote{Note that we take Ethereum’s “node count” (https://www.ethernodes.org/) and not its “validator count” (https://ethereum.org/en/staking/), the latter of which does not adequately reflect the network’s number of validator nodes.}
    \item The original measurements included by Platt et al \cite{Platt2021TheProof-of-Work} are replicated and used, but also other, newer, empirical observations. As a result, the number of observations per \ac{DLT} grows to up to 7 instead of 2 and covers a longer timeframe.
    \item With a greater number of observations available, we do not just extrapolate an affine function from two points but conduct a linear regression to describe the association between throughput and the number of validators.
    \item An additional assumption is introduced that with a throughput of zero, the number of validators is also zero. This is included as an observation at t0 in the aforementioned regression.
    \item The same methodology is employed to estimate Bitcoin’s energy consumption, but the Cambridge Centre for Alternative Finance’s data is updated \cite{CBECI2023CambridgeIndex}.
    \item The same methodology is employed to estimate VisaNet’s energy consumption, but the latest Visa energy report \cite{VISA20212021Report} is used.
    \item The assumption that throughput and the number of validators are positively correlated is maintained.
    \item The lower and upper bounds for the power demand of a validator are improved and updated where possible by resorting to empirical measurements for each \ac{DLT} \cite[15]{Gallersdorfer2022EnergyProtocols,CCRI2022DeterminingNetworks} \cite[23]{CCRI2022TheNetwork}, except where they contradict official estimates \cite{SolanaFoundation2022Solanas2022,SolanaFoundation2022Validator2022}. Where new data is unavailable, the bounds are assumed to be the same as in \cite{Platt2021TheProof-of-Work}. Table III (Appendix) specifies both bounds for each \ac{DLT}.
    \item The L1 maximum throughput figures are reviewed through a bibliography review via online search engines. Where maximum recorded throughput exceeds our sources for maximum throughput estimates,\footnote{E.g. https://metrics.algorand.org/\#/protocol/ } the former records are preferred. This resulted in an upward revision of Tezos’ and Algorand’s maximum throughput and a downward revision of Ethereum’s maximum throughput (before sharding).
\end{enumerate}

\section{Methodology}
\label{sec:methodology}

The first step in our methodology is to observe the number of validators ($\quad N_{\text {val }}$) and throughput (l) for a given \ac{DLT} at a given moment in time (see dates in Table VIII, Appendix). A function extrapolating the relationship between the two variables is built therefrom, specifically with $\quad N_{\text {val }}$ as a function of l. The $\quad N_{\text {val }}$ variable is then translated to an energy consumption value using different assumptions on the possible electricity usage per validator. By introducing the number of transactions as a denominator, the energy consumption “per transaction” is arrived at.

The energy consumption of a given validator, called  p, depends on the hardware used. The product between  p and $\quad N_{\text {val }}$ is the energy cost of the network with that specific configuration of hardware and number of validators. If we divide this energy consumption by the number of transactions in the corresponding period, we obtain the consumption per transaction \cite{Platt2021TheProof-of-Work}.

Energy consumption per transaction is however only comparable at the same level of throughput, with comparisons that exclude this denominator being potentially misleading. For this reason, energy consumption is modelled as a function of throughput, whereby the number of validators depends on the number of transactions. The hypothesis, which Platt et al \cite{Platt2021TheProof-of-Work} find to have some empirical support and hence plausibility, is that as more users join a network, “of the new users, a share becomes validators and another non-disjoint share executes transactions, meaning that $\quad N_{\text {val }}$ and l are positively correlated” \cite[1137]{Platt2021TheProof-of-Work}.

\begin{equation} \label{eq:model}
f c_{t x}(l)=\frac{\left(N_{v a l} \times p\right)}{l} 
\end{equation}

\begin{align*}
\text{where} \quad N_{\text {val }} = \text{Number of validators} \\
\text{and} \quad p = \text{Energy consumption per validator} \\
\text{and} \quad l = \text{Transactions per second (throughput)}
\end{align*}

Although the precise information of the hardware (and, hence, the energy consumption) of each validator is not available, we seek to improve upon the optimistic and pessimistic energy consumption thresholds proposed by Platt et al \cite{Platt2021TheProof-of-Work} as described in section Replication and expansion.

As a result, we obtain a function describing the $\quad N_{\text {val }}$ of each \ac{DLT} for a given level of l. The number of validators is then translated to a p range using the aforementioned thresholds. The domain of the function is defined by the energy consumption at the maximum postulated throughput without layer 2 solutions, which can be seen in Table VII (Appendix).

Platt et al \cite{Platt2021TheProof-of-Work} proposed an affine function relating the number of validators on the \ac{DLT} and its throughput. We do the same, and assume it is valid to use a linear regression to adjust the following model:

\begin{equation}
N_{\text {val }}=k \times \lambda l
\end{equation}

This gives the next expression:

\begin{equation}
f c_{t x}(l)=\frac{(k \times \lambda l) \times p}{l}
\end{equation}

For Platt et al \cite{Platt2021TheProof-of-Work} two values were derived for affining $\mathrm{k}$ and $\lambda$, resulting in Table \ref{tab:tableplatt}.

\begin{table}[htp]
\centering
\begin{tabular}{lrr}
\hline 
\textbf{Platform} & $\mathrm{k}$ & $\lambda$ \\ 
\hline 
Algorand & 102.8 & 103.9 \\ 
Cardano & 1267.8 & 2959.2 \\ 
Polkadot & 297 & 0 \\ 
Tezos & 440.7 & -24.6 \\ 
Hedera & 7.6 & 0.3 \\ 
\hline
\end{tabular}
\caption{Estimates for $\mathrm{k}$ and $\lambda$ for different \ac{DLT} platforms used in (3) to model the number of validators depending on the number of transactions per second, Table III Platt et al \cite[1138]{Platt2021TheProof-of-Work}.}
\label{tab:tableplatt}
\end{table}

Our new values for $\mathrm{k}$ and $\lambda$ can be observed in Figure 2 (section Results).

\subsection{Limitations}
\label{subsec:limitations}

As a replication study, this article brings forward some of the limitations\footnote{Another limitation outlined by the authors, namely that “while our model suggests that \ac{PoS} systems can remain energy-efficient while scaling up to VisaNet throughput levels, there is no hard evidence in support of this argument, as, to our knowledge, no \ac{DLT}-based system has experienced a sustained volume of this magnitude to date on the base level,” has been mitigated with the relatively higher throughput levels observed in Hedera and Solana in 2022 and/or early 2023.} already outlined by Platt et al \cite[1137-1140]{Platt2021TheProof-of-Work}:

\begin{quote}
“We therefore only consider validators (...) The overall number of nodes, including other full nodes that replicate the transaction history without participating in consensus, is likely higher for all systems analysed.”
\end{quote}

\begin{quote}
“We have used broad consumption ranges to model the energy consumption of individual validator nodes. While we are confident that the actual energy consumption is in fact within these ranges, the underlying characteristics of different \ac{PoS} protocols that might impact energy consumption, such as the accounting model, have been ignored.”
\end{quote}

\begin{quote}
“While assuming that the electricity consumption of a validator node is independent of system throughput is well justified for the permissionless systems analysed [32], permissioned systems that are designed to support high throughput may not warrant such an assumption.”
\end{quote}

\begin{quote}
“We have not so far distinguished between transaction types.”\footnote{Exploring this further is an area for Future work. For instance, the site Metrika (https://app.metrika.co/) distinguishes multiple transaction types for Algorand (Stpf, Keyreg, Afrz, Acfg, Appl, Pay, Axfer) and Hedera (Crypto Create Account, HCS, HTS - Fungible Tokens, HTS - NFTs, Smart Contract, File).}
\end{quote}

\begin{quote}
“We ignored the possibility of achieving effectively higher throughput than the specified maximum through layer 2 (L2) solutions, such as the Lightning Network or via optimistic rollups and zero-knowledge (zk)-rollups that are receiving increasing attention.”
\end{quote}

We may furthermore identify the following limitations:

\paragraph{Univariable model and predictive power}
This paper models  $\quad N_{\text {val }}$ as a function of a single variable, namely l. In doing so, it does not seek to achieve maximal predictive power. We are aware that many factors may influence $\quad N_{\text {val }}$ and do not seek to construct a complex and comprehensive model to predict the evolution of this variable. Rather, we undertake an exploratory task by attempting to conduct a first approximation to meaningful energy consumption comparisons, understanding that it is essential to control for l (and that it is plausible to expect throughput to impact the number of validators).

A high R2 is thus not expected for all \ac{DLT}s. A low R2 should be considered both a legitimate and natural result, but also a limitation on the ability to predict based on this model.

\paragraph{Linearity}
This paper assumes both the relationship
\begin{enumerate}
    \item between $\quad N_{\text {val }}$ and l and
    \item between $\quad N_{\text {val }}$ and p
\end{enumerate}
to be linear. This replicates the methodology established in Platt et al \cite{Platt2021TheProof-of-Work}. This is assumed because we find it plausible that:
as more users join a network to execute transactions, a share of them becomes validators \cite{Platt2021TheProof-of-Work};
the main source of a node’s electricity consumption is its idle consumption and replicated execution, which (unlike the number of edges between nodes, i.e. network complexity) evolve linearly with $\quad N_{\text {val }}$.
However, if:
\begin{enumerate}
    \item the evolution of $\quad N_{\text {val }}$ depending on l could be better explained as a polynomial function; and/or
    \item there were significant performance/efficiency degradation with network scale/complexity within the function domain,\footnote{See also Sedlmeir et al (2021).}
\end{enumerate}
this could have implications for our findings. In the first case, there would be a need for proper curve-fitting and interpretation, requiring an in-depth evaluation of the characteristics of each \ac{DLT}. In the second case, our model would unfairly benefit \ac{DLT}s that are relatively smaller at the time of observation, and should be amended to consider that energy consumption evolves quadratically with the number of validators. Nevertheless, we assume that performance stays within reasonable boundaries for the range of $\quad N_{\text {val }}$ values under study (i.e. $\quad N_{\text {val }}$ values corresponding to l values lower within the domain established by Table VII in the Appendix).

Our findings (see Results) already suggest that, for some local periods in the short run, some of the \ac{DLT}s under study do not display a linear relationship between $\quad N_{\text {val }}$ and l. For the period studied, for instance, Polkadot maintained a fixed number of 297 validators, meaning that l rose and fell independently of $\quad N_{\text {val }}$. Hedera also maintained fixed numbers of validators, which it periodically increased with the addition of new Council members; in some periods, a fall in l meant there were negative relationships between $\quad N_{\text {val }}$ and l. However, and even if this may reduce the R2, we believe that the function positively associating $\quad N_{\text {val }}$ and l plausibly holds in the long term, as a scenario where throughput falls and yet more and more validators join the network seems unsustainable in the long run. This would suggest that the energy consumption of these two \ac{DLT}s might be overestimated in this paper.

\paragraph{Data availability}
The methodology used to compute the validator count may differ across the sources used in this paper (Table IV, Appendix). For instance, Solana Foundation’s (2022c)\footnote{https://solanabeach.io/} validator count includes “delinquent” consensus nodes which other sources do not consider as current nodes.\footnote{https://solscan.io/validator} To avoid a slippery slope, we take validator count figures at face value. Data availability also means that different sources are used to compute the ranges for validator power consumption for Binance Smart Chain (Opentaps, 2022 and Gallersdörfer et al, 2022), Elrond/MultiversX, Flow, Hedera, NEAR and Toncoin \cite{Platt2021TheProof-of-Work}, Ethereum \cite{CCRI2022TheNetwork}, Solana \cite{Gallersdorfer2022EnergyProtocols,SolanaFoundation2022Solanas2022,SolanaFoundation2022Validator2022} and the other \ac{DLT}s \cite{Gallersdorfer2022EnergyProtocols}. Future work should obtain comparable real-world data for all \ac{DLT}s.

Sometimes available data sources differed and we were unable to ascertain the reason therefor. Specifically, for Polkadot, Polkastats\footnote{https://polkastats.io/} throughput data differed from Subscan\footnote{https://polkadot.subscan.io/transfer} data; we resorted to the former for consistency as Polkastats also provides validator count data. However, when past values displayed at Polkastats were not retrievable through web archive search engines, we used Subscan. On that note, it should be highlighted that, unlike throughput data (for which both current and past data was usually available), historic validator count data was rarely available, which hampered data collection. We resorted to web archives where needed and possible to mitigate this issue.

\paragraph{Observation period and observation randomness}
Due to manpower limitations, the dates of empirical observations of the number of validator nodes and throughput are not distributed homogeneously. Nevertheless, observation dates were randomly chosen in the sense that no specific pattern was involved in their selection other than data availability and the availability of the research team.

Because of this, and our experience of a marginally decreasing usefulness of each additional data point introduced, we are confident that the introduction of additional measurements within the observation period will not substantially change the results of this paper. Nevertheless, we encourage replication.

However, the observation window is in itself a limitation indeed. It barely exceeds one year, which we deem as insufficient to ideally derive a function describing the evolution of $\quad N_{\text {val }}$ with l, especially considering that in the short run many other factors may affect the evolution of $\quad N_{\text {val }}$ and l (see Discussion). For this reason, this paper will continue to be updated with additional observations following the release of the working paper, to improve the density and quality of the data.

Additionally, the reader should note that, first, observations for Ethereum cover only a fraction of the observation period for other \ac{DLT}s, as Ethereum is only considered in the post-Merge period. Second, some \ac{DLT}s (BNB Chain, Elrond/MultiversX, Flow, NEAR, Toncoin and Tron) were only included in a more recent iteration of our research, with data collection thus starting later for these networks as well.

\paragraph{Data comparability}
In the year 2022, one of the networks studied – Solana – suffered several public relations scandals (Nambiampurath,  2022), one of which impacted the source data used in this article. Specifically, it became known that Solana’s most widely circulated throughput figures include a combination of both “vote” and “nonvote” transactions, the majority of which were the former. While it is technically true that vote transactions are transactions in Solana, many argued that they should be excluded from Solana’s \ac{TPS} count, as vote transactions are a part of the consensus mechanism, and not “real” transactions in the meaningful sense that enables comparison with \ac{DLT}s (SolanaFM, 2022; Zhu, 2022). To address this issue, we switched to a different data source that counts only nonvote transactions (https://solana.fm/). For observations executed before the scandal, we apportioned Solana’s tps in the proportion of nonvote transactions to total (vote + nonvote) transactions for the observation day according to https://analytics.solscan.io/ (see Table V in Appendix). For Solana’s maximum throughput, we divided Solana’s maximum postulated throughput in official sources by the average ratio of nonvote transactions to total (vote + nonvote) transactions.

\paragraph{Energy consumption “per transaction”}
Our methodology does not account for transaction complexity, and relies on an unrealistic assumption that any given transaction is equal to any other in terms of both value and energy consumption, or that the distribution of transaction complexity is comparable across \ac{DLT}s. Future work should seek to incorporate this nuance in its modelling and observations.

Furthermore, the reader should note that the measurement of energy transaction on a “per-transaction” basis has in itself been criticised, particularly when used to benchmark against Bitcoin Core (see \cite{Carter2022BitcoinZero,Saylor2022BitcoinAgency}). The “per-transaction” basis is used in this paper as it constitutes the standard proxy for energy intensiveness used to in the field for comparison purposes, and its usage for comparison purposes in non-mining settings has some plausibility, as transactions may be considered a driver of energy consumption\footnote{See above the assumption that throughput drives adoption by user, which drives $\quad N_{\text {val }}$.} in PoS-based \ac{DLT}s.
However, usage of this metric should not be taken to mean that the number of transactions processed is a direct and proximate cause of energy consumption, especially for PoW systems. Bitcoin’s energy consumption is almost entirely a subproduct of mining, which in turn depends on Bitcoin’s price, that only depends on the number of transactions in a very indirect manner \cite{Saylor2022BitcoinAgency}.

Most importantly, Bitcoin advocates are concerned that the “per-transaction” metric is misused to suggest that if Bitcoin already consumes an already significant amount of energy with low throughput, it will need to consume an even greater amount of energy to replace higher throughput systems like VisaNet. This is not the case because (a) throughput is not the cause of Bitcoin’s electricity consumption; (b) Bitcoin’s L1 throughput is enshrined in its parameters, i.e. it is a protocol limitation by design; (C) therefore, Bitcoin could scale arbitrarily without increasing its energy usage by changing its parameters or adopting L2 solutions such as the Lightning Network, meaning it may even scale its “value settled” arbitrarily without scaling its energy consumption \cite{Saylor2022BitcoinAgency}.

We do not make any of these claims and moreover consider some of them partially applicable to PoS-based \ac{DLT}s as well. In this context, the “per-transaction” metric is introduced as a useful proxy for comparison purposes, which has value within these limitations.

\paragraph{Trade-offs and fairness}
This paper does not consider other variables such as network security, fault-tolerance, ordering fairness, access fairness, timestamping fairness, decentralisation of state maintenance,\footnote{This includes the relationship between decentralisation and membership size, see Vergne \cite{Vergne2020DecentralizedPlatform}.} decentralisation of digital asset distribution, risk of capture, censorship-resistance, etc. \cite{Madsen2019FairExplained,Hedera2020HederaTransactoins,Hedera2020HederaTransactoins,RaphaelBiner2022FairThesis,Saylor2022BitcoinAgency}. Consequently, the trade-offs between these variables and energy consumption are outside the scope of this paper; any comparison assuming otherwise would not constitute a fair assessment.\footnote{In an extreme scenario, a naïve application of this paper’s methodology could be misused to make a broken, unfair and unsafe system “look good” because of its very low energy consumption. The reader should note, however, that although unfairly misused, stating the fact of the broken system’s low energy consumption would nonetheless be “true” and energy consumption is the only variable within the scope of this paper.}

In addition, there is a sense of efficiency whereby what matters is the (throughput-controlled) energy consumption per transaction “per validator:” algorithm efficiency. This paper does not use this metric because the number of validators is already controlled by modelling many possible values of $\quad N_{\text {val }}$, and because energy consumption is a proxy of environmental impact,\footnote{Insofar \ac{GHG} emissions are assumed to be the same, which may not be the case \cite{Ibanez2023CanSoK}.} a variable that is interesting as a whole (total energy consumption) or in terms of cost-benefit (energy consumption per transaction) but not directly in terms of how many validators participated in its generation (energy consumption per transaction per validator). Nevertheless, learning the efficiency of the \ac{PoS} algorithm itself is of indirect environmental value and, shall comparable testing conditions be built (see Future work and \cite{Gallersdorfer2022EnergyProtocols}, their investigation should be encouraged.

Despite these limitations, we are confident that this article accurately replicates, expands, and updates the methodology proposed by \cite{Platt2021TheProof-of-Work} and that this constitutes a substantial contribution to the state of the art.

\section{Results}
\label{sec:results}

\begin{figure*}[h!]
\centering
\includegraphics[width=\textwidth]{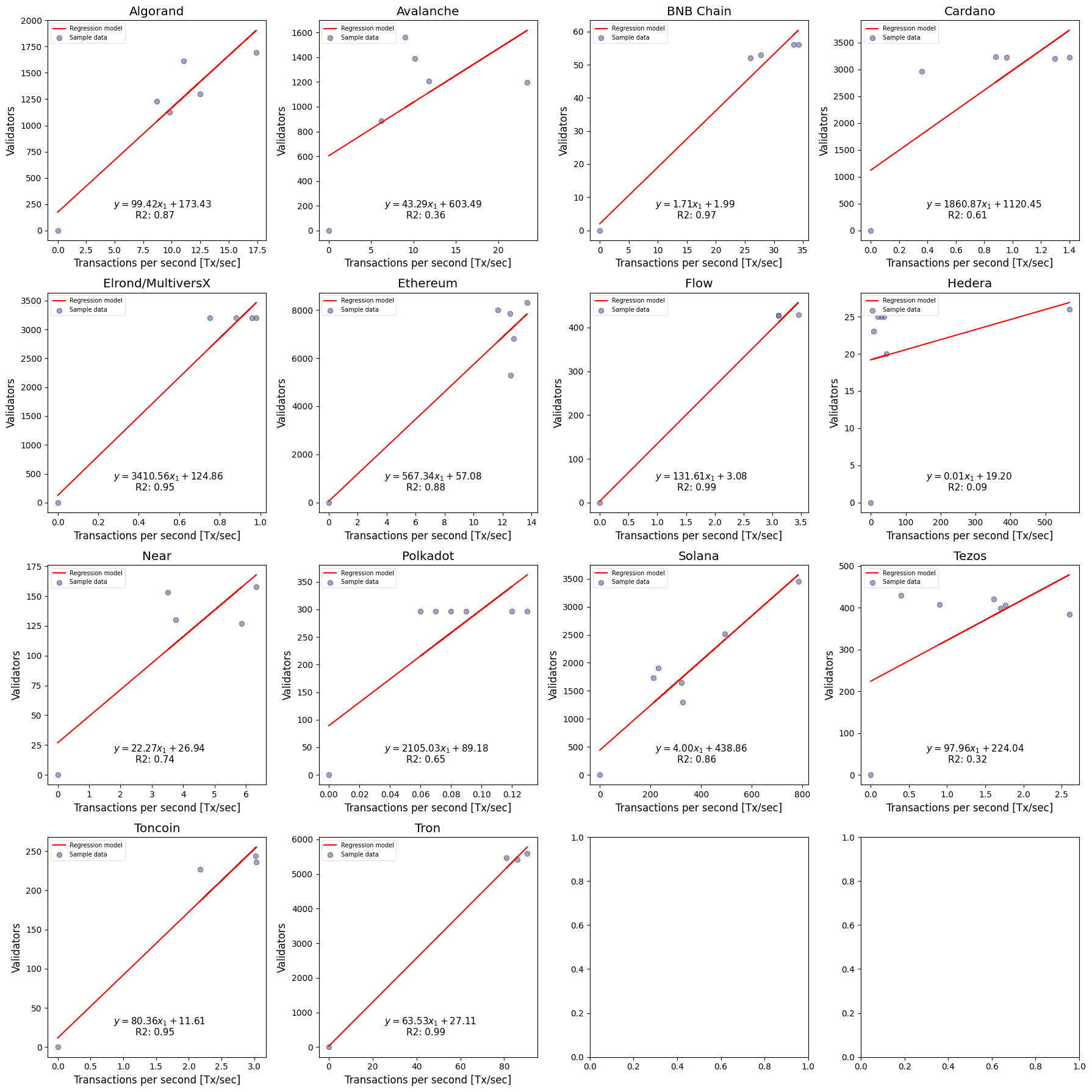}
\caption{Replication of function estimates for $f c_{t x}(l)$, i.e. update of table III of Platt et al (2021: 1138).}
\label{figure:regression}
\end{figure*}

The replication and update of the paper provided the outputs (global and per transaction energy consumption estimates) that can be observed in Table VIII (Appendix). With these measures we re-adjust the $\mathrm{k}$ and $\lambda$, resulting in Figure \ref{figure:regression}.

\begin{figure*}[h!]
\centering
\includegraphics[width=\textwidth]{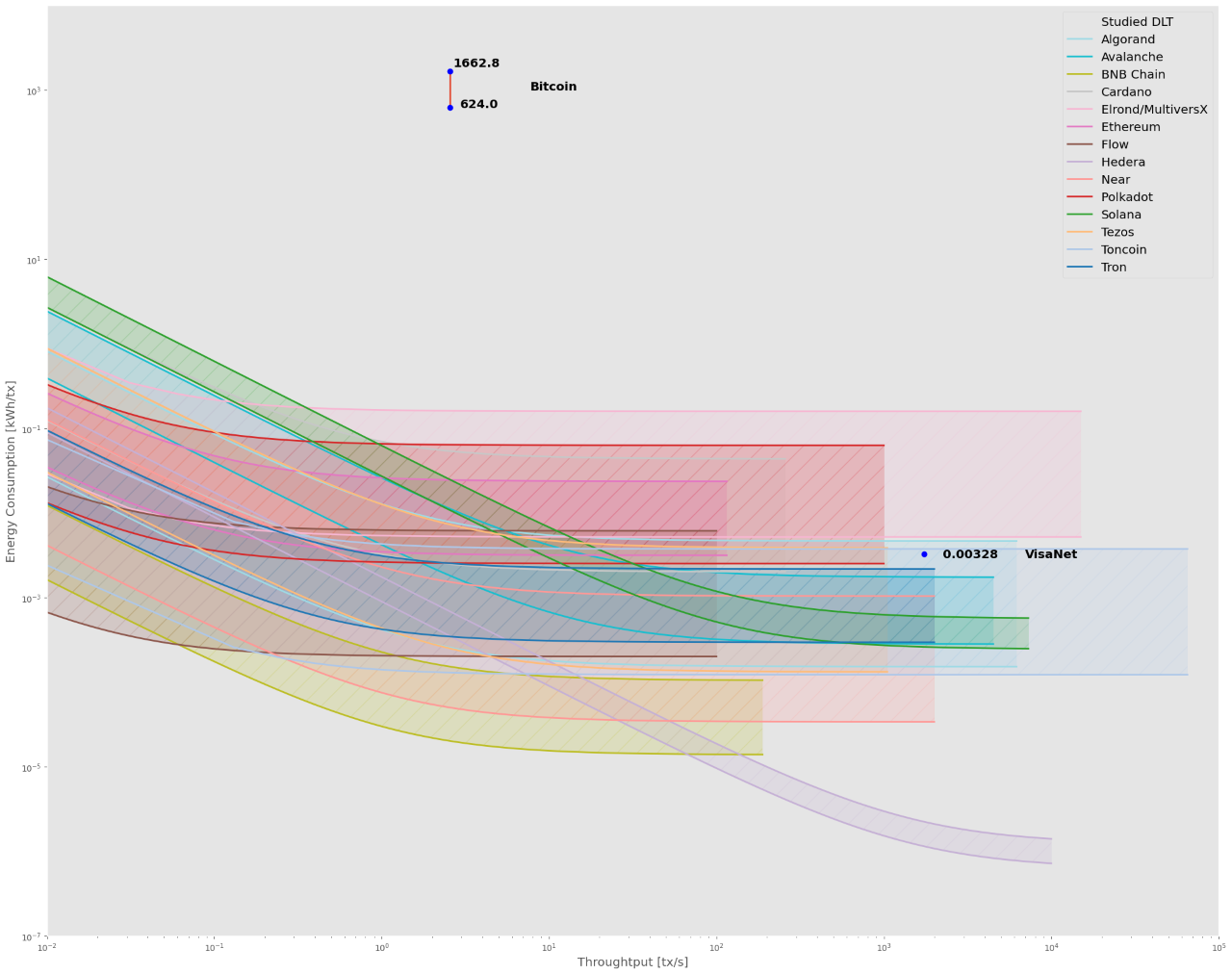}
\caption{Extrapolation of throughput-controlled energy consumption per transaction. For each system’s range, the lower boundary indicates the energy consumption under an optimistic validator hardware assumption while the upper threshold indicates a pessimistic scenario. For readability, see this chart broken down into two in Figures 4 and 5 (Appendix).}
\label{figure:mainresults}
\end{figure*}

These new values of $\mathrm{k}$ and $\lambda$ were used to update the extrapolation chart (Figure \ref{figure:mainresults}).

\subsection{Discussion}
\label{subsec:discussion}

With these findings, we reaffirm the usefulness of the energy consumption estimation methodology introduced in Platt et al (2021), and suggest improvements. We confirm that the energy consumption of \ac{PoS}-based \ac{DLT} systems is several orders of magnitude lower than that of Bitcoin Core (see also \cite{Gallersdorfer2022EnergyProtocols}), and the potential combinations of throughput and energy consumption for \ac{PoS}-based \ac{DLT}s are on par, if not superior, to VisaNet in terms of energy efficiency.

We observe that “The Merge” led to a significant decrease in Ethereum’s energy consumption. Ethereum’s latest global power consumption is estimated to range 963 kW to 33 kW, whereas its energy consumption per transaction is estimated between 0.02095 kWh and 0.0007188724 kWh, which compares to the rangles expected by Platt et al \cite[1139]{Platt2021TheProof-of-Work}. This means that the consumption for Ethereum may have fallen more than anticipated: 445.3 to 14.6 and 0.00803 to 0.00026, respectively.

We observe that not all PoS-based \ac{DLT}s are born equal, and that this statement holds in various different dimensions:
\begin{enumerate}
    \item In terms of their absolute energy consumption.
    \item In terms of their energy consumption per transaction.
    \item In terms of their energy consumption per transaction for a given level of throughput.
    \item In terms of the shape of the function describing the evolution of their energy consumption per transaction with throughput
\end{enumerate}

The progressive addition of additional data points also allowed us to observe that after the fifth data point the changes in the resulting functions were only marginal. This is not to be interpreted as disregarding the need for more data: strongly to the contrary, more data would be highly beneficial to this model and its improvement. However, it suggests that additional data should be incorporated \textit{over a longer period of time}. In contrast, the addition of further data points in the time period already studied is not expected to materially affect the findings.

We also observe that the R2 of the functions describing the correlation of throughput with the number of validators varies significantly across \ac{DLT}s. This implies that variation in the energy consumption of the \ac{DLT}s under study may be caused by many factors other than network throughput, and that this paper only uncovers a part of this variable's behavior. We hold thus that this paper produces a comparison of throughput-controlled energy consumption that is valid and useful, but also one that shows limitations as a predictor and should not be treated as such.

Specific observations about particular \ac{DLT} networks may be formulated. We note that Solana’s energy consumption per transaction (relative to its peers) is substantial at low throughput; however, it is estimated to be significantly more efficient, in relative terms, at high levels of throughput.

We may also note that the curve describing Polkadot’s energy consumption per transaction is slightly flatter than its peers’ and than itself in Platt et al \cite{Platt2021TheProof-of-Work}. This may be attributed to the current 297 validator cap that was in place during the entire observation period for both this paper and Platt et al’s. In the latter, global energy consumption was estimated to be constant, with the energy consumption per transaction steadily decreasing with throughput increases, as a result. In our paper, the assumption of zero throughput with zero validators changes this, producing a different function to describe Polkadot’s behavior. However, as long as the validator cap is in place, the values of this function may be safely assumed to change substantially with the number of observations in the same period, because the intercept will have a greater “weight” in the computation of the function with a smaller number of observations. For reasons explained at the beginning of this paper, we believe our assumption to result in a more plausible function than Platt et al’s. Nevertheless, the utility of the function remains very limited and only effectuating additional observations in the long run\footnote{In this context, we define the long run as a time period long enough such that it exceeds the period in which the 297 validator cap is in place.} may improve it.

Both in this study and in Platt et al \cite{Platt2021TheProof-of-Work}, Cardano is consistently shown to be one of the \ac{DLT}s with the highest energy consumption in relative terms (despite being negligible in all cases). Although further research would be required to confirm this, this may be due the high number of validators and comparatively low throughput that it displayed throughout the entire observation period. This could be hypothesised to be related to Cardano’s usage of the UTXO model \cite{Platt2021TheProof-of-Work} at a low block size.

We have noted a series of Limitations to our methodology and encourage the reader to familiarise with them, as well as Future work to be focused in addressing them.

To continue improving the work presented in this article, data collection will continue after the release of this working paper. Furthermore, the source data and methodology is made available for free at the Online Appendix, to encourage feedback, replication and expansion of the work presented in this article.

Finally, we note that it is of interest to the blockchain ecosystem to obtain simple figures comparing the energy consumption per transaction and the throughput and validator value for the various \ac{DLT}s. If we take the latest observation (“contemporary” observation in Platt et al \cite{Platt2021TheProof-of-Work}) and average out lower and upper bounds for energy consumption, the following results:

\begin{table*}[h!]
\centering
\label{tab:dlts}
\begin{tabular}{ccccc}
\hline \ac{DLT} & Global power consumption $(\mathrm{kW})$ & Energy consumption/tx $(\mathrm{kWh})$ & TPS & Validators \\
\hline Algorand & $106.82$ & $0.003411$ & $8.70$ & 1,227 \\
Avalanche & $101.62$ & $0.002395$ & $11.79$ & 1,209 \\
Bitcoin & $92,325,000,000.00$ & 2,927 & $2.56$ & $10,000+$ \\
BNB Chain & $7.02$ & $0.000059$ & $33.4$ & 56 \\
Cardano & $142.63$ & $0.041270$ & $0.96$ & 1,209 \\
Elrond/ & & & & \\
MultiversX & $277.76$ & $0.102875$ & $0.75$ & 3,200 \\
Ethereum & $450.15$ & $0.009956$ & $12.56$ & 5,294 \\
Flow & $37.15$ & $0.003318$ & $3.11$ & 428 \\
Hedera & $6.45$ & $0.000003$ & $568.45$ & 26 \\
Near & $13.71$ & $0.000602$ & $6.33$ & 158 \\
Polkadot & $16.66$ & $0.035593$ & $0.13$ & 297 \\
Solana & $917.29$ & $0.000517$ & $493.00$ & 2,512 \\
Tezos & $29.81$ & $0.009203$ & $0.90$ & 407 \\
Toncoin & $21.18$ & $0.001948$ & $3.02$ & 244 \\
Tron & $391.92$ & $0.001202$ & $33.40$ & 56 \\
VisaNet & $1,736.00$ & $0.003280$ & 1,736 & n/a \\
\hline
\end{tabular}
\caption{Estimates for global and per-transaction energy consumption (average between upper and lower bound) and values for throughput and number of validators, in the last observation only (see Figure \ref{figure:mainresults} for the overall trend).}
\end{table*}

However, we encourage the reader to use Figure \ref{figure:mainresults} instead, as energy consumption per transaction may be a misleading metric when the throughput that underlies the observation differs significantly.

\subsection{Future work}
\label{subsec:future}

As established also in Platt et al \cite{Platt2021TheProof-of-Work} (see also \cite{CCRI2022DeterminingNetworks}), future work should consider factors such as transaction complexity, the energy consumption of non-validating nodes (e.g. auxiliary nodes), comparable empirical measurements for the actual hardware configurations in each \ac{DLT}, and the effects of moving from a permissioned to a permissionless \ac{DLT} setting. Future research may also expand even further the universe of \ac{PoS}-based \ac{DLT}s studied, or collect additional observations in a more systematic manner. 

We furthermore encourage researchers to undertake the complex task of building a more comprehensive model of $\quad N_{\text {val }}$, depending not just on l, to achieve proper predictive capacity of not just $\quad N_{\text {val }}$ itself but also energy consumption. There is also much potential for additional work that does consider L2 solutions, as this paper assumes exclusively L1 technology.

Future work may also construct meaningful comparisons of (throughput-controlled) energy consumption per transaction “per validator,” to better comprehend the nuances across different \ac{PoS} algorithms, inefficiencies in implementation, etc. This should not be achieved by merely dividing energy consumption by the number of validators, however, but with physical measurements, by building test networks to impose comparable throughput levels, networking requirements and hardware requirements (see \cite{Gallersdorfer2022EnergyProtocols}).

As anticipated earlier, there is also potential for future work whereby the goal is more ambitious, namely the prediction of energy consumption under different values for throughput. This would require curve-fitting, which is beyond the scope of our paper.

Finally, and as anticipated in the Limitations section, future research may explore nonlinear relationships between variables, efficiency degradation with scale, trade-offs of energy efficiency with other variables (such as decentralisation, security, fault-tolerance, fairness), and nuances with validator count figures across different sources, as well as both extend the observation period and introduce additional (random) observations within it.

\section{Conclusion}

With this article, we show that the model proposed by Platt et al \cite{Platt2021TheProof-of-Work} for measuring the energy consumption of a PoS-based \ac{DLT} is replicable. We suggest improvements and furthermore expand, refine and update the model with additional \ac{DLT}s, data and assumptions. Overall, we confirm the core finding of Platt et al \cite{Platt2021TheProof-of-Work} that, regardless of the nuances and differences across PoS-based \ac{DLT}s,\footnote{See also \cite{CCRI2022EnergyBlockchain}.} all of the PoS-based \ac{DLT}s analysed have an energy consumption that is negligible compared to that of major \ac{PoW} blockchains. To the extent that energy consumption may be considered problematic, this is not an issue in any \ac{PoS} design.

We invite the scientific community to further expand this work, and encourage \ac{DLT} foundations to share with us further data to iterate this model.

\section*{Acknowledgements}
\label{sec:acknowledgements}

We thank Jiahua Xu and Paolo Tasca for their invitation and support to write this entry. We thank Emin Gün Sirer for the comments provided to the co-authors of Platt et al \cite{Platt2021TheProof-of-Work} which enriched this article as well.

%\section*{Author Contributions}
%...

\section*{Conflict of Interest}

Juan Ignacio Ibañez was one of the original co-authors in the paper by Platt et al \cite{Platt2021TheProof-of-Work}. Both Juan Ignacio Ibañez and Francisco Rua are affiliated to the UCL Centre for Blockchain Technologies (UCL CBT), which maintains professional and institutional relationships with organisations linked to several of the \ac{DLT}s under review.

\printacronyms

\printbibliography

%\newpage

\raggedbottom

\vspace{1em}

%\appendix

\onecolumn

\section*{Appendix}

\begin{table}[htbp]
\centering
\begin{tabular}{ccp{3cm}cp{3cm}}
\hline
\hline \ac{DLT} & Lower bound (W) & Source & Upper bound (W) & Source \\
\hline Algorand & 5.53 & Measurement in lab setting (Gallersdörfer et al, 2022: 15) & 168.59 & Measurement in lab setting (Gallersdörfer et al, 2022: 15) \\
Avalanche & 23.44 & Measurement in lab setting (Gallersdörfer et al, 2022: 15) & 144.67 & Measurement in lab setting (Gallersdörfer et al, 2022: 15) \\
BNB Chain & 29.25 & Minimum possible value in Gallersörfer et al (2022: 15) for the hardware options estimated by Opentaps (2022) & 221.33 & Maximum possible value in Gallersörfer et al (2022: 15) for the hardware options estimated by Opentaps (2022) \\
Cardano & 3.90 & Measurement in lab setting (Gallersdörfer et al, 2022: 15) & 84.47 & Measurement in lab setting (Gallersdörfer et al, 2022: 15) \\
Elrond/MultiversX & 5.50 & Estimate of typical power consumption for a small singleboard computer Raspberry Pi 4 (Platt et al, 2021: 1138) & 168.10 & Estimate of typical power consumption for a general-purpose rackmount server Dell PowerEdge R730 (Platt et al, 2021: 1138) \\
Ethereum & 20.00 & Measurement in lab setting (CCRI, 2022b: 23) & 150.06 & Measurement in lab setting (CCRI, 2022b: 23) \\
Flow & 5.50 & Estimate of typical power consumption for a small singleboard computer Raspberry Pi 4 (Platt et al, 2021: 1138) & 168.10 & Estimate of typical power consumption for a general-purpose rackmount server Dell PowerEdge R730 (Platt et al, 2021: 1138) \\
Hedera & 168.10 & Estimate of typical power consumption for a general-purpose rackmount server Dell PowerEdge R730 (Platt et al, 2021: 1138) & 328.00 & Power consumption under 80\% load for a high-performance server Hewlett Packard Enterprise ProLiant ML350 Gen10 (Platt et al, 2021: 1138) \\
NEAR & 5.50 & Estimate of typical power consumption for a small singleboard computer Raspberry Pi 4 (Platt et al, 2021: 1138) & 168.10 & Estimate of typical power consumption for a general-purpose rackmount server Dell PowerEdge R730 (Platt et al, 2021: 1138) \\
Polkadot & 4.31 & Measurement in lab setting (Gallersdörfer et al, 2022: 15) & 107.86 & Measurement in lab setting (Gallersdörfer et al, 2022: 15) \\
Solana & 221.33 & Measurement in lab setting (Gallersdörfer et al, 2022: 15) & 509.00 & Official average power consumption estimation (Solana Foundation, 2022a) \\
Tezos & 4.86 & Measurement in lab setting (Gallersdörfer et al, 2022: 15) & 141.65 & Measurement in lab setting (Gallersdörfer et al, 2022: 15) \\
Toncoin & 5.50 & Estimate of typical power consumption for a small singleboard computer Raspberry Pi 4 (Platt et al, 2021: 1138) & 168.10 & Estimate of typical power consumption for a general-purpose rackmount server Dell PowerEdge R730 (Platt et al, 2021: 1138) \\
Tron & 16.80 & Measurement in lab setting (CCRI, 2022a: 13) & 123.50 & Measurement in lab setting (CCRI, 2022a: 13) \\
\hline
\end{tabular}
\label{tab:bounds}
\caption{Conceivable upper and lower bounds for the power consumption of the average validator of each PoS-based \ac{DLT} under study.}
\end{table}

\begin{table*}
\centering
\begin{adjustbox}{width=\textwidth}
\begin{tabular}{|l|l|l|l|}
\hline
\textbf{Platform} & \textbf{Validators} & \textbf{Throughput} & \textbf{\ac{TPS} source variable} \\
\hline
Algorand & \url{https://algoexplorer.io} & \url{https://metrics.algorand.org/} & Transactions per second (last 24 hours) \\
Avalanche & \multicolumn{2}{|c|}{\url{https://subnets.avax.network/}} & Transactions per day \\
BNB Chain & \url{https://bscscan.com/validators} & \url{https://bsctrace.com/} & Transactions per second (last 5 minutes) \\
Cardano & \multicolumn{2}{|c|}{\url{https://cardanoscan.io/}} & Transactions per day \\
Elrond & \url{https://multiversx.com/} & \url{https://explorer.elrond.com/} & Transactions per day \\
Ethereum & \url{https://www.ethernodes.org/history} & \url{https://etherscan.io/chart/tx} & Transactions per day \\
Flow & \url{https://flowscan.org/staking/overview} & \url{https://flowscan.org/metrics/blocks} & Transactions per day \\
Hedera & \url{https://status.hedera.com/} & \url{https://hedera.com/dashboard} & Transactions per day \\
NEAR & \multicolumn{2}{|c|}{\url{https://explorer.near.org/}} & Transactions per day \\
Polkadot & \url{https://polkastats.io/es} & \url{https://polkadot.subscan.io/extrinsic}; & Transactions per day \\ &  &  \url{https://polkastats.io/es} &  \\
Solana & Solana Foundation (2022c), \url{https://solscan.io/validator} & \url{https://analytics.solscan.io/}, \url{https://solana.fm/} & Transactions per hour; Transactions per second \\
Tezos & \url{https://tzstats.com/bakers} & \url{https://tzstats.com/} & Transactions per day \\
Toncoin & \url{https://ton.org/} & \url{https://tonapi.io/} & Transactions over 60 seconds (empirical observation) \\
Tron & \url{https://tronscan.org/#/blockchain/nodes} & \url{https://tronscan.org/#/blockchain/transactions} & Transactions per day \\
\hline
\textbf{Platform} & \textbf{Energy consumption} & \textbf{Throughput} & \textbf{Variable definition} \\
\hline
\textbf{Bitcoin} & \url{https://ccaf.io/cbeci/index} & \url{https://www.blockchain.com/explorer} & \ac{TPS} (last day) - \\ & &  & Annual energy consumption in \ac{GJ}\\
        \textbf{Visa} & \multicolumn{2}{|c|}{https://usa.visa.com/content/dam/VCOM/}\\
& \multicolumn{2}{|c|}{regional/na/us/about-visa/documents/2021-}\\
& \multicolumn{2}{|c|}{environmental-social-and-report.pdf} & \ac{TPS} (not specified) - Annual energy consumption in TWh\\
      \hline
\hline
\end{tabular}
      \end{adjustbox}
        \label{tab:sources}
        \caption{Sources for the data used. Where unavailable, past snapshots from Wayback Machine were used.}
\end{table*}

\begin{table}
\centering
\begin{tabular}{|c|c|c|c|c|c|c|c|}
\hline
\textbf{Platform} & \textbf{Time period} & \multicolumn{3}{|c|}{\textbf{Annual energy consumption}} & \textbf{Energy per second} & \textbf{Tx/s} & \textbf{Energy per tx} \\
 &  & \multicolumn{2}{|c|}{\textbf{Source measure}} & \textbf{kWh} & \textbf{kWh} & &  \textbf{kWh/tx} \\
\hline
Visa & 2021 & 646,000 & GJ & 179,439,420.00 & 5.69 & 1,736 & 0.00328 \\
Bitcoin lower & 2022 & 50.41 & TWh & 50,410,000,000.00 & 1,598.49 & 2.56 & 624.0 \\
Bitcoin upper & 2022 & 134.24 & TWh & 134,240,000,000.00 & 4,256.72 & 2.56 & 1662.8 \\
\hline
\end{tabular}
\caption{Estimation of energy consumption of Visa and Bitcoin (lower and upper bounds).}
\label{tab:energy-consumption-comparison}
\end{table}

\begin{table}[htbp]
\centering
\begin{tabular}{ccccc}
\hline \ac{DLT} & Global power consumption $(\mathrm{kW})$ & Energy consumption/tx $(\mathrm{kWh})$ & TPS & Validators \\
\hline Algorand & $106.82$ & $0.003411$ & $8.70$ & 1,227 \\
Avalanche & $101.62$ & $0.002395$ & $11.79$ & 1,209 \\
Bitcoin & $92,325,000,000.00$ & 2,927 & $2.56$ & $10,000+$ \\
BNB Chain & $7.02$ & $0.000059$ & $33.4$ & 56 \\
Cardano & $142.63$ & $0.041270$ & $0.96$ & 1,209 \\
Elrond/ & & & & \\
MultiversX & $277.76$ & $0.102875$ & $0.75$ & 3,200 \\
Ethereum & $450.15$ & $0.009956$ & $12.56$ & 5,294 \\
Flow & $37.15$ & $0.003318$ & $3.11$ & 428 \\
Hedera & $6.45$ & $0.000003$ & $568.45$ & 26 \\
Near & $13.71$ & $0.000602$ & $6.33$ & 158 \\
Polkadot & $16.66$ & $0.035593$ & $0.13$ & 297 \\
Solana & $917.29$ & $0.000517$ & $493.00$ & 2,512 \\
Tezos & $29.81$ & $0.009203$ & $0.90$ & 407 \\
Toncoin & $21.18$ & $0.001948$ & $3.02$ & 244 \\
Tron & $391.92$ & $0.001202$ & $33.40$ & 56 \\
VisaNet & $1,736.00$ & $0.003280$ & 1,736 & n/a \\
\hline
\end{tabular}
\label{tab:power}
\caption{Contemporary observations per \ac{DLT} studied.}
\end{table}

\begin{table*}[htbp]
\centering
\resizebox{\textwidth}{!}{%
\begin{tabular}{ccccccc}
\hline $\begin{gathered}\text { \textbf{Tx/s} } \\
\text { \textbf{observation} }\end{gathered}$ & \textbf{Date} & \textbf{Nonvote tx per day} & \textbf{Total tx per day} & \textbf{Average tx/s} & $\begin{gathered}\text { \textbf{Nonvote vs} } \\
\text { \textbf{total tx ratio} }\end{gathered}$ & \textbf{Nonvote tx/s} \\
\hline 1734 & $14 / 9 / 21$ & $31,436,549$ & $166,730,469$ & 1,930 & $0.189$ & 327 \\
2227 & $30 / 3 / 22$ & $26,040,310$ & $179,416,101$ & 2,077 & $0.145$ & 323 \\
2020 & $2 / 5 / 22$ & $18,166,816$ & $174,322,626$ & 2,018 & $0.104$ & 211 \\
3363. & $28 / 7 / 22$ & $36,691,080$ & $157,490,743$ & 1,823 & $0.233$ & \\
3338 & $9 / 26 / 22$ & $35,338,176$ & $185,904,800$ & 2,152 & $0.190$ & 635 \\
3183 & $11 / 10 / 22$ & $17,855,155$ & $170,894,804$ & 1,978 & $0.104$ & 333 \\
4123 & $11 / 12 / 22$ & $172,633,381$ & $309,222,640$ & 3,579 & $0.056$ & 230 \\
\hline
\end{tabular}
}
\label{tab:solana}
\caption{Calculation of past Solana nonvote throughput based on https://analytics.solscan.io/ and https://solana.fm/.}
\end{table*}

\begin{table}[htbp]
\centering
\begin{tabular}{ccp{3cm}cp{3cm}}
\hline
\textbf{\ac{DLT}} & \textbf{Source} & \textbf{Metric} & \textbf{Value} \\
\hline
Algorand & \url{https://metrics.algorand.org/#/protocol/} & Peak TPS & 6,192 \\
Avalanche & \url{https://support.avax.network/en/articles/5325146-what-is-transactional-throughput} & Transactions per second & 4,500 \\
BNB Chain & \url{https://bscscan.com/chart/tx} & Maximum recorded transactions per second & 188 \\
Cardano & \url{https://vacuumlabs.com/what-we-love-about-cardano-a-technical-analysis/} & Transactions per second & 257 \\
Elrond & \url{https://docs.elrond.com/welcome/welcome-to-elrond/} & Transactions per second & 15,000 \\
Ethereum & \url{https://ethtps.info/} & Maximum recorded transactions per second & 115 \\
Flow & \url{https://developers.flow.com/flow/faq/operators} & Transactions per second & 100 \\
Hedera & \url{https://hedera.com/hbar} & Transactions per second & 10,000 \\
NEAR & \url{https://near.org/papers/the-official-near-white-paper/} & Transactions per second & 2,000 \\
Polkadot & \url{https://twitter.com/gavofyork/status/1255859146127179782} & Transactions per second & 1,000 \\
Solana & \url{https://twitter.com/solana/status/1334565204449841153} & Transactions per second * nonvote tx to total transaction ratio & 7,295 \\
Tezos & \url{https://forum.tezosagora.org/t/tezos-storage-irmin-march-2022/4497} & Transactions per second & 1,048 \\
Toncoin & \url{https://coinmarketcap.com/currencies/toncoin/} & Maximum recorded transactions per second & 65,000 \\
Tron & \url{https://tron.network/static/doc/white\_paper\_v\_2\_0.pdf} & Transactions per second & 2,000 \\
\hline
\end{tabular}
\label{tab:maxtps}
\caption{Maximum throughput per \ac{DLT} under study.}
\end{table}

\end{document}